\documentclass{article}
\usepackage{spconf,amsmath,graphicx}
\usepackage{graphicx}
\usepackage{amsmath, amssymb, bm}
\usepackage{subfigure}
\usepackage{epstopdf}
\usepackage{amsthm}
\usepackage{cases}
\usepackage{cite}
\usepackage{enumitem}
\usepackage{mathtools}
\usepackage{multirow}
\usepackage{bbm}
\usepackage{color}
\usepackage{ascmac}
\usepackage{arydshln}
\usepackage[table]{xcolor}
\usepackage{float}
\usepackage{fancybox}
\usepackage{mdframed}
\usepackage{hyperref}
\usepackage{bbm}

\newtheorem{definition}{Definition}

\graphicspath{{fig/}}

\newcommand{\cv}{\mathbf{c}}

\newcommand{\fv}{\mathbf{f}}

\newcommand{\uv}{\mathbf{u}}

\newcommand{\xv}{\mathbf{x}}
\newcommand{\yv}{\mathbf{y}}

\newcommand{\Am}{\mathbf{A}}
\newcommand{\Bm}{\mathbf{B}}

\newcommand{\Dm}{\mathbf{D}}

\newcommand{\Id}{\mathbf{I}}

\newcommand{\Lm}{\mathbf{L}}

\newcommand{\Um}{\mathbf{U}}
\newcommand{\Wm}{\mathbf{W}}

\newcommand{\Ec}{\mathcal{E}}

\newcommand{\Gc}{\mathcal{G}}

\newcommand{\Vc}{\mathcal{V}}

\newcommand{\Lambdam}{\hbox{\boldmath$\Lambda$}}

\title{LOSSY COMPRESSION OF ADJACENCY MATRICES BY GRAPH FILTER BANKS}
\name{
Kenta Yanagiya$^{1}$ \qquad Junya Hara$^{1}$ \qquad Hiroshi Higashi$^{1}$ \qquad Yuichi Tanaka$^{1}$ \qquad Antonio Ortega$^{2}$
\thanks{This work was supported in part by JSPS KAKENHI under Grant 23H01415, JST SPRING under Grant JPMJSP2138, and JST AdCORP under Grant JPMJKB2307.}
}
\address{
$^{1}$ Osaka University, Osaka, Japan\\
$^{2}$ University of Southern California, Los Angeles, CA, USA
}
\begin{document}
\ninept
\maketitle
\begin{abstract}
This paper proposes a compression framework for adjacency matrices of weighted graphs based on graph filter banks.
Adjacency matrices are widely used mathematical representations of graphs and are used in various applications in signal processing, machine learning, and data mining. 
In many problems of interest, these adjacency matrices can be large, so efficient compression methods are crucial.
In this paper, we propose a lossy compression of weighted adjacency matrices, 
where the binary adjacency information is encoded losslessly (so the topological information of the graph is preserved) while the edge weights are compressed lossily.
For the edge weight compression, the target graph is converted into a \textit{line graph}, whose nodes correspond to the edges of the original graph, and where the original edge weights are regarded as a graph signal on the line graph.
We then transform the edge weights on the line graph with a graph filter bank for sparse representation.
Experiments on synthetic data validate the effectiveness of the proposed method by comparing it with existing lossy matrix compression methods. 
\end{abstract}
\section{Introduction}
\label{sec:intro}
Graphs are mathematical representations that have many applications to analyze social, brain, sensor, and transportation networks, to name a few \cite{ortega2018, shuman2013, tanaka2018, tanaka2020, yamada2020, sakiyama2019a, onuki2016}.
Despite their utility, graphs with a large number of nodes and edges demand a lot of computational resources for storing and transmitting graph information.
Therefore, effective graph compression methods are required.

There exist many lossless graph compression schemes for unweighted graphs.
Representative examples include conversions to bitmap, $k^2$ trees, and succinct data structures \cite{besta2019, alvarez2010, khandelwal2017, martinez-bazan2012}.
Lossless compression of unweighted graphs is generally efficient (and may be reaching its theoretical performance limits \cite{besta2018}) and desirable (since topological information can be fully preserved). 
However, efficient lossless compression of weighted graphs is difficult unless they have some special properties, such as sparsity.
Therefore, lossy compression is often used for weighted graphs.
In this paper, we propose to focus on \textit{lossy graph compression for weighted graphs}, where the topology (the corresponding unweighted graph) is coded losslessly.  

Lossy graph compression considers the trade-off between reconstruction quality and compression ratio.
Many lossy graph compression methods have been proposed, including node and edge sampling, graph sparsification, and graph coarsening \cite{sakiyama2019a, yanagiya2022, spielman2011a, jin2020}, which are typically based on \textit{graph simplification}, i.e., some unnecessary characteristics are discarded while keeping some important ones.

In weighted graphs, edges convey information about (1) the existence of interconnections between nodes (edge on/off) and (2)  the strength of edges (edge weights). 
Most lossy compression approaches for weighted graphs consider the edge existence and edge weights simultaneously during simplification \cite{willcock2006, zhou2012, toivonen2011, liu2012}, reducing the number of nodes and edges. The performance of these methods can be quantified as a function of errors in edge weights and degree distribution. These methods modify the original graph topology, which is generally undesirable, as they can greatly affect graph-based applications and visualization of graphs.
These changes also modify the spectrum of the graph and thus can have a significant impact on various graph signal processing (GSP) methods.
Therefore, the compression should consider not only simple edge weight errors but also the impact on methods that use the spectrum of the graph, such as signal diffusion on the graph and spectral clustering.
Our work is motivated by the assumption that topological information is more important than edge weights: this leads to the idea of compressing them separately \cite{besta2019, jin2020}.

We propose a two-step compression method: The adjacency information describing the presence of edges is losslessly compressed,  while edge weights are compressed in a lossy manner.
Since the adjacency information is represented as a binary matrix, we can use existing lossless compression methods for unweighted graphs, such as those described above.
Thus, we focus on lossy compression of the edge weights. We propose a method that takes into account the adjacency information, which is available at both the encoder and the decoder (since it is transmitted lossesly).
In our approach, the original adjacency information is first converted into a \textit{line graph}, whose nodes and edges represent the edges of the original graph and their relationships, respectively \cite{aigner1967, evans2010}.
With this conversion, vectorized edge weights are regarded as a \textit{graph signal} on the line graph.
Therefore, we can efficiently compress the edge weights with GSP techniques \cite{narang2012, sakiyama2019, narang2013}.
This leads to a compression of edge weights that considers the relationships among edges implicit in the graph.

We show experimentally that the proposed method can improve compression by over $25\%$ compared to existing methods that achieve the same reconstruction error.
We also conduct clustering experiments showing that existing lossy compression methods are unstable regardless of the compression ratio while the proposed method, which can preserve topology, always shows high consistency between clustering results.
\begin{figure*}[t]
    \centering
    \includegraphics[width = 0.75\linewidth]{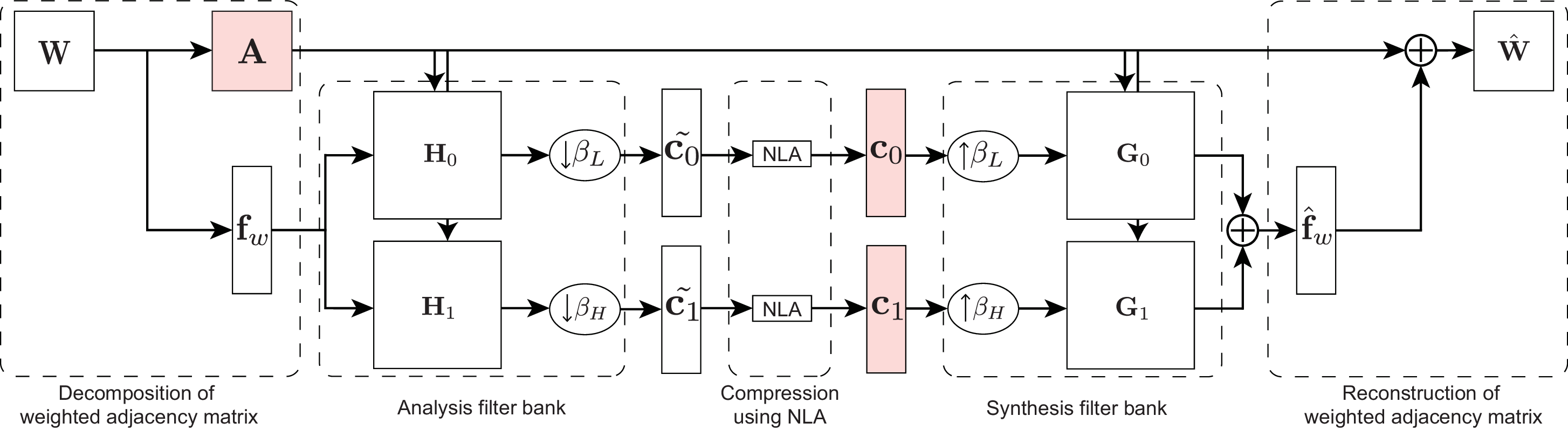}
    \caption{Overall framework of the proposed adjacency matrix compression method. The matrix and vector that are compressed and stored are highlighted in red.}
    \label{fig:Overview_Compression}
\end{figure*}

\section{Related Work}
\textbf{Lossless and lossy matrix compression} methods can be used for graphs since graphs can be represented as matrices.
\textit{Lossless} matrix compression techniques typically employ entropy coding, such as arithmetic coding and Huffman coding \cite{hu2000, besta2019}.
For sparse matrices, sparse matrix representation that preserves only nonzero elements, such as compressed sparse column (CSC) data structures \cite{willcock2006}, may effectively reduce storage usage.
\textit{Lossy} methods include using the Fourier transform or the singular value decomposition to project the matrix onto a subspace \cite{pavlo2009, distasi2006} and then quantizing or thresholding the resulting matrix spectrum, which typically results in a low-rank approximation of matrices.

\textbf{Graph compression} can be viewed as a type of matrix compression.  
The main approaches include 
graph simplification, which reduces the number of edges \cite{sakiyama2019a, spielman2011a, yanagiya2022}, graph summarization and aggregation, which creates a homogeneous graph by estimating super-nodes and super-edges \cite{khan2015}. 
Note that these methods do not have bitrate scalability: If we need to obtain a different bitrate, it is required to rerun the algorithm to achieve the required sparsity. 

\textbf{Graph filter banks} are one of the techniques used to compress graph signals \cite{narang2012, narang2013, sakiyama2014, sakiyama2019}.
A typical graph filter bank comprises analysis and synthesis transforms, where each of the transforms includes a set of graph filters and sampling operations.
Therefore, appropriate selections of graph filters and sampling methods are key for graph signal compression.

\section{LOSSY COMPRESSION OF WEIGHTED ADJACENCY MATRIX}

\subsection{Preliminaries}
An undirected weighted graph is represented as $\Gc = (\Vc, \Ec, \Wm)$, where $\Vc$ and $\Ec$ are the sets of nodes and edges, respectively, and $\Wm$ is the weighted adjacency matrix, where the $(i,j)$-entry $w_{ij}$ represents the edge weights between node $i$ and node $j$.
$N = |\Vc|$ is the size of node set.
The degree matrix $\Dm$ is diagonal, and its $i$th element $[\Dm]_{ii} = d_i = \sum_j [\Wm]_{ij}$ refers to the degree of node $i$.
Given the weighted adjacency matrix and degree matrix, we can define the graph Laplacian matrix $\Lm = \Dm - \Wm$.
Since $\Lm$ is a real symmetric matrix, it always has a complete set of orthonormal eigenvectors $\uv_i$.
$\uv_i$ has a non-negative real eigenvalue $\lambda$ satisfying $\Lm\uv_i = \lambda \uv_i$.
Then, we can write the eigendecomposition of $\Lm$ as $\Lm = \Um\bm{\Lambda}\Um^\top$.
$\Um$ is a orthonormal matrix with $\uv_i$ aligned, and $\bm{\lambda} = \text{diag}(\lambda_0, \cdots, \lambda_{N-1})$.
These eigenvalues are ordered as $0=\lambda_{0}<\lambda_{1} \leq \lambda_{2} \leq \cdots \leq \lambda_{N-1}=\lambda_{\max}$.

In this paper, we use both directed and undirected incidence matrices.
The directed incidence matrix $\Bm$ is defined as: 
\begin{equation}\label{eqn:directed_incidence_matrix}
    [\Bm]_{i\alpha} = 
    \begin{cases}
        1 & \text{Edge } \alpha=(i,j) \textrm{ is incident to node } i,\\
        -1 & \text{Edge } \alpha=(i,j) \textrm{ is incident to node } j,\\
        0 & \text{Otherwise}.
    \end{cases}
\end{equation}
Directed incidence matrices can also be defined for undirected graphs by considering pseudo-orientation.
For the line graph, we also define the undirected incidence matrix as:
\begin{equation}\label{eqn:undirected_incidence_matrix}
    [\tilde{\Bm}]_{i\alpha} = \begin{cases}
    1 & \text{Edge } \alpha=(i,j) \textrm{ is incident to node } i,\\
    0 & \text{Otherwise}.
\end{cases}
\end{equation}

\subsection{Framework}
An overview of the proposed weighted adjacency matrix compression is shown in Fig.~\ref{fig:Overview_Compression}.
We losslessly transmit the binary adjacency information of the graph and lossily compress the edge weights using a graph filter bank.
The proposed method has the advantages that 1) topological information such as connectivity is preserved, and 2) edge weight compression can be performed with bitrate scalability.

In the proposed method, a given weighted adjacency matrix $\Wm$ is separated into a binary adjacency matrix $\Am \in \{0,1\}^{N\times N}$ and an edge weight vector $\fv_w \in \mathbb{R}_{\ge 0}^{|\Ec|}$, where 
$\Am$ is given by:
\begin{equation}
    [\Am]_{ij} = \begin{cases}
        1 & [\Wm]_{ij} \neq 0,\\
        0 & \text{otherwise.}
    \end{cases}
\end{equation}
One of the existing lossless compression methods can then be applied to compress $\Am$.

As for the weights signal, there is no obvious way to order the elements in $\fv_w$ since the indexing of the edges is arbitrary (unlike the indices of samples in a time series).
Therefore, while $\fv_w$ is a one-dimensional vector, techniques for regular domain signals (e.g., wavelet transform for regular signals) do not result in an efficient compression.

To compress $\fv_w$ efficiently, we propose using the graph topology information available at both the encoder and decoder since we losslessly compress $\Am$.
Note that the original graph represented by $\Am$ captures the inter-node relationships, but we need the inter-edge relationships to compress $\fv_w$.
Thus, we construct the \textit{line graph} of $\Am$, $\Gc_L$, whose nodes represent the edges of the original graph and whose edges represent relationships between the edges of the original graph \cite{evans2010}. 
We consider the edge weight vector, $\fv_w$, as a graph signal on $\Gc_L$, so edge weights sharing a common node in $\Gc$ are signal values on neighboring nodes in $\Gc_L$.
Note that edge weights in line graphs often tend to be smooth, especially when nodes are placed evenly in space \cite{yanagiya2022}.
This allows us to exploit correlations between the edge weights in $\Gc$ using a line-graph filter bank (LGFB) to obtain a signal decomposition of $\fv_w$. 
The proposed method transforms $\fv_w$ into subband coefficients $\tilde{\cv}_k$ ($k=0,\dots, M-1$ where $M$ is the number of channels) by a graph filter bank.
The transformed coefficients are then compressed.

The main difference between our proposed method and existing techniques for weighted graphs\cite{akritas2004, willcock2006, zhou2012, toivonen2011, liu2012} is that we employ separate compression methods for the topology (lossless) and the edge weights (lossy, with bitrate scalability.)

\begin{figure}[t]
    \centering
    \includegraphics[width=\linewidth]{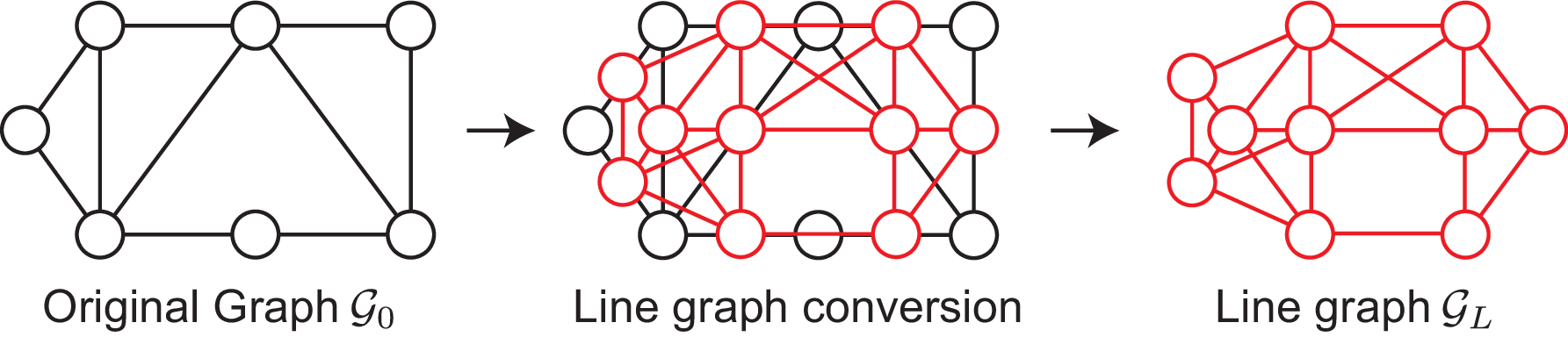}
    \caption{Illustration of line graph conversion. The red node represents an edge of the original graph. An edge of a line graph is drawn if the two edges are connected via a node in the source graph.}
    \label{fig:line_graph_conversion}
\end{figure}

\subsection{Line Graph Conversion and Graph Operator for Inter-Edge Relationships}\label{sec:line graph}
This section defines the line graph and graph operators for the edge domain.
Line graph conversion is defined as follows.

\begin{definition}[Line graph]
\label{def:Line Graph}
Suppose an undirected incidence matrix $\tilde{\Bm}$ \eqref{eqn:undirected_incidence_matrix} associated with the original graph $\Gc = (\Vc, \Ec)$ is given.
The adjacency matrix $\Am_L \in \mathbb{R}^{|\Ec| \times |\Ec|}$ of $\Gc_L$ is \cite{evans2010}:
\begin{equation}
\Am_L = \tilde{\Bm}^\top\Bm - 2\Id_{|\Ec|}
\label{eqn:adj_line}
\end{equation}
where $\alpha$ and $\beta$ are edge indices of the original graph (and therefore, they are node indices in $\Gc_L$)
and $\Id_{|\Ec|} := \text{diag}(\mathbf{1})$.
\end{definition}
\noindent
The line graph conversion process is illustrated in Fig.~\ref{fig:line_graph_conversion}.
In addition to the graph Laplacian of $\Gc_L$ induced from the adjacency matrix of \eqref{eqn:adj_line}, 
we also use the \textit{edge Laplacian} as a possible operator.
\begin{definition}[Edge Laplacian]
\label{def:Edge laplacian}
Suppose a directed incidence matrix $\Bm$ \eqref{eqn:directed_incidence_matrix} corresponding to the original graph $\Gc$ is given.
The edge Laplacian is defined as follows \cite{roddenberry2022}:
\begin{equation}
    \Lm_e = \Bm^\top\Bm.
\end{equation}
The edge Laplacian is the Hodge Laplacian with dimension $2$, while 
the graph Laplacian of the original graph is the Hodge Laplacian of dimension $1$ \cite{roddenberry2022}. 
\end{definition}

\subsection{Graph Filter Banks on Line Graphs (LGFBs)}
As discussed above, we can use any graph filter bank to transform $\fv_w$.
Two conditions need to be considered when selecting a filter bank. 
First, many graph filter banks assume that the underlying graph is bipartite \cite{narang2012, narang2013}. However,  
line graphs are not bipartite in general, and therefore, graph filter banks that can be used on arbitrary graphs are preferable.

Second, the computational cost of filtering and sampling should be manageable. 
As the number of channels in a graph filter bank increases, the computation time required for filtering and sampling increases accordingly. 
Thus, while the sparseness of the transformed coefficients may be promoted when filterbanks with many channels are used, this comes at the cost of an increase in overall computation costs. 
The number of edges in the original graph is generally larger than the number of nodes, i.e., $|\Ec| \gg N$.
This implies that the LGFB should be critically sampled with few channels.

\begin{figure}
    \centering
    \includegraphics[width=\linewidth]{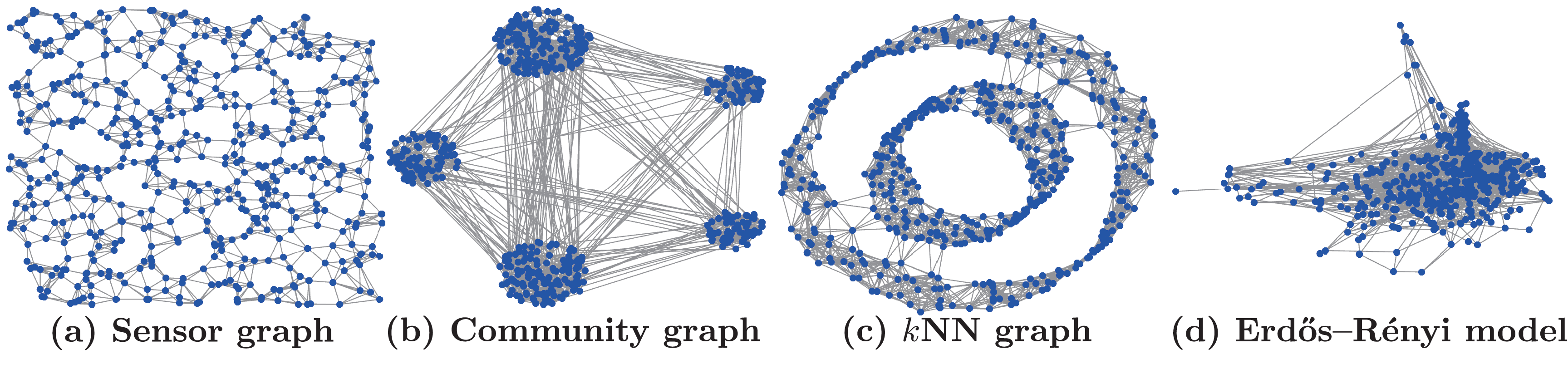}
    \caption{Examples of graphs with $N = 500$. (b) and (c) are graphs with 5 and 2 clusters, respectively.}
\label{fig:graph}
\end{figure}

\section{Experimental Results}
\subsection{Setup}
In our experiments, we use the following weighted graphs with $N  = 500$: 1) Random sensor graph: $|\Ec| = 1,781$, 2) Community graph with five communities: $|\Ec| = 4,792$, 3) $k$NN graph constructed from point clouds ($k = 15$): $|\Ec| = 4,188$, and 4) Random graph based on Erd\H{o}s--R\'{e}nyi model with a node connection probability $0.05$: $|\Ec| = 3,097$.
Fig. \ref{fig:graph} shows examples of the graphs.
Edge weights are randomly drawn from a truncated normal distribution where the mean is set to $1$ and the truncated range is $[0, 2]$.

In the proposed method, we use GraphBior \cite{narang2013} as the graph filter bank.
While our method applies to any graphs, GraphBior (and similar graph filter banks) requires  bipartition of the graph:
We apply the Harary decomposition \cite{narang2012} for graph bipartition.
In future work, we plan to consider newly designed filter banks for arbitrary graphs \cite{pavez2023}. 
Transformed coefficients are sparsified via nonlinear approximation (NLA) (refer to Fig. \ref{fig:Overview_Compression}).
The proposed methods using graph Laplacian of $\Gc_L$ and edge Laplacian (see Section \ref{sec:line graph}) are abbreviated as \texttt{Proposed-line} and \texttt{Proposed-edge}, respectively.

In contrast to our proposed compression methods, existing methods directly compress weighted adjacency matrices.
We compare the performance with the following approaches.

\noindent
\textbf{Transformation with graph filter banks} (\texttt{Direct-GFB}): Since GFBs can transform graph signals, this approach transforms columns of a weighted adjacency matrix sequentially. NLA is applied to the 
transformed coefficients, as in our proposed method.

\noindent
\textbf{Low-rank approximation} (\texttt{Direct-LRA}) \cite{akritas2004}: In direct compression of the adjacency matrix through LRA, the singular value decomposition (SVD) of $\Wm$ is computed, and the smaller singular values are set to zero.

\noindent
\textbf{Transformation with DCT} (\texttt{Direct-DCT}): In this approach, $\Wm$ is transformed with DCT in a column-wise manner, and the transformed coefficients are thresholded with NLA.

\noindent
\textbf{Binary adjacent information} (\texttt{Binary}): To clarify the effect of edge weights, we also compare the results only using the binary adjacency matrix, i.e., no weight information is transformed or transmitted.

The thresholded coefficients of all methods are compressed with uniform quantization and Huffman coding.
The quantization step was experimentally set to $0.01$.

\begin{figure}[t]
\centering
\subfigure[][Sensor graph]{\includegraphics[width=0.48\linewidth]{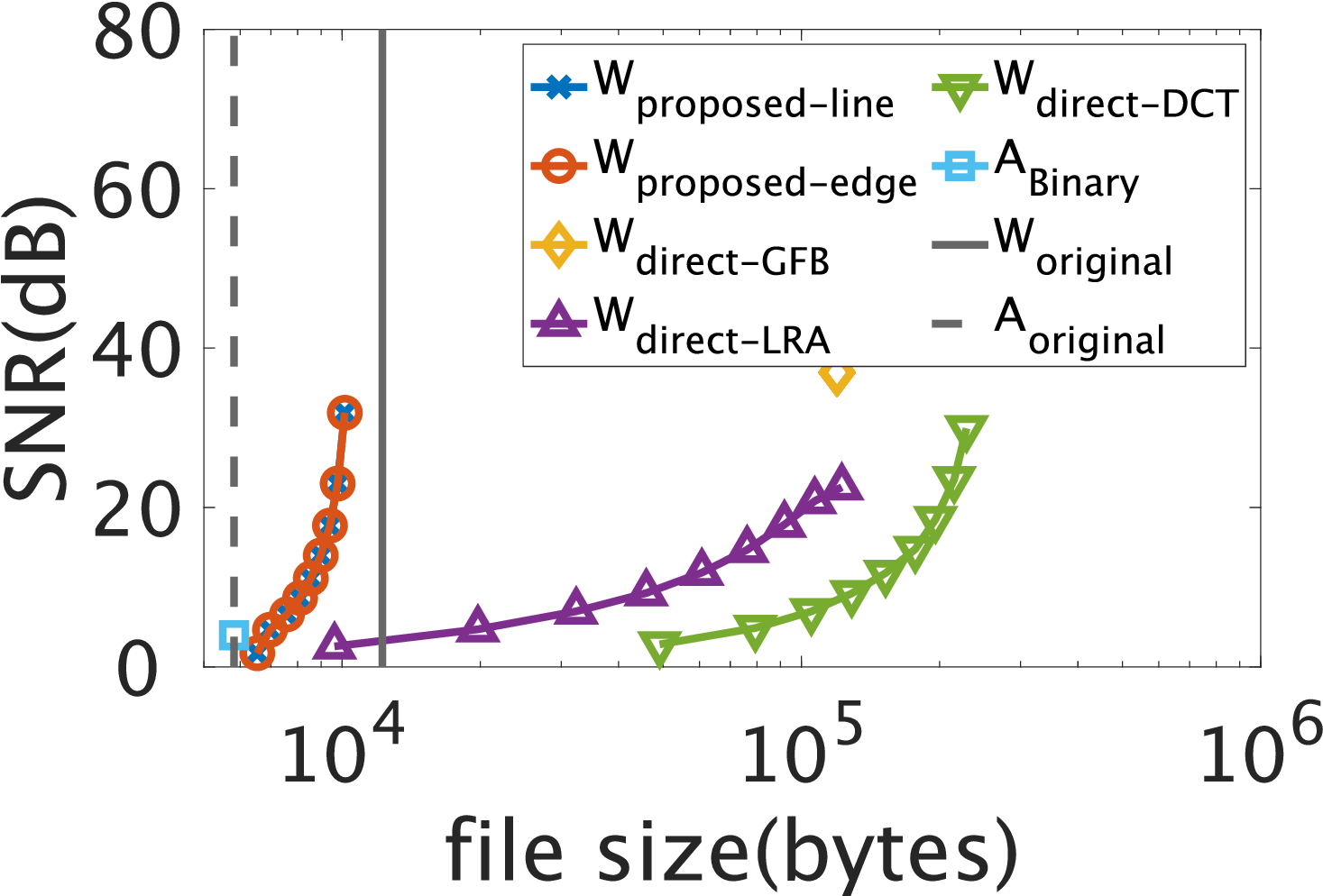}}
\subfigure[][Community graph]{\includegraphics[width=0.48\linewidth]{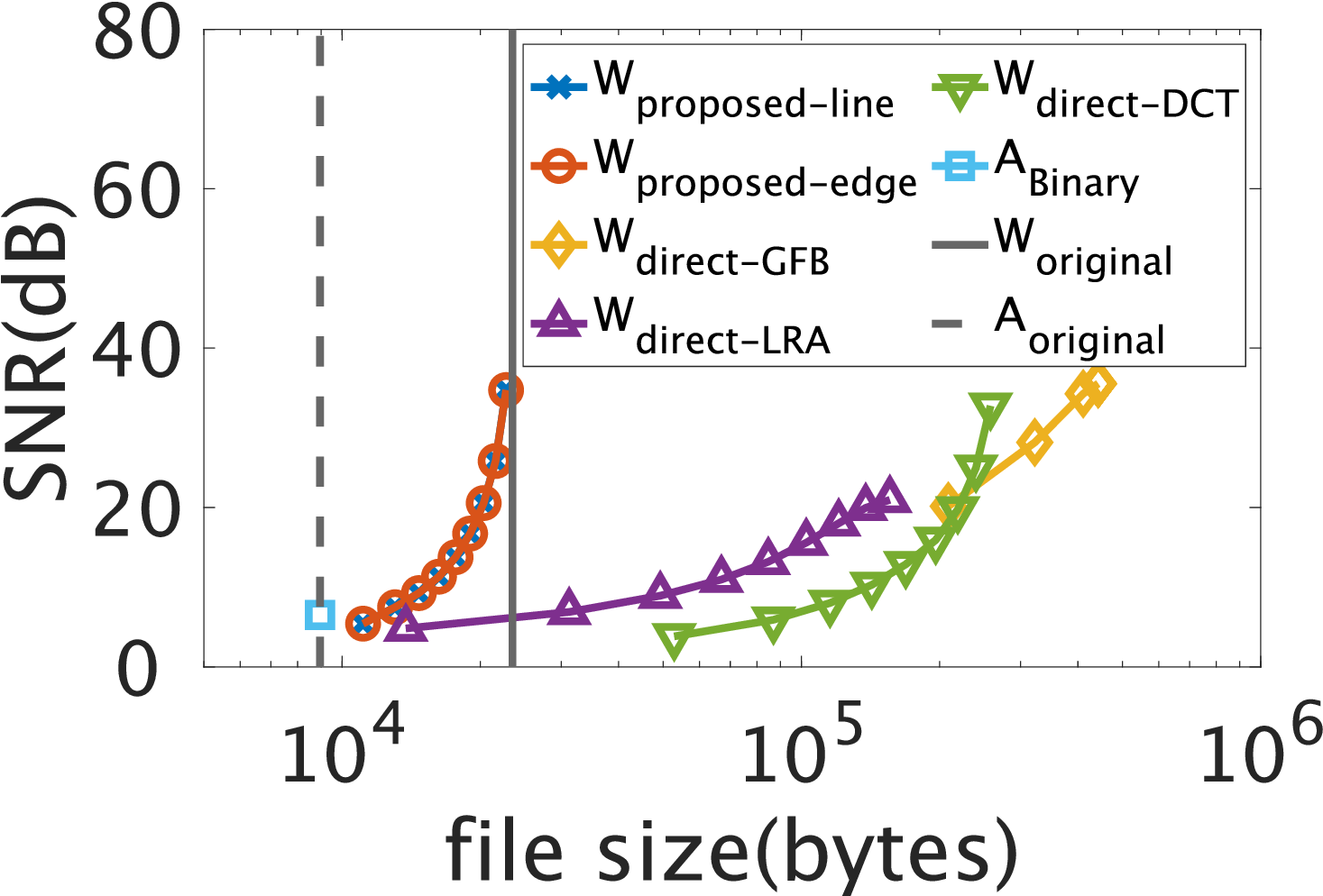}}\\
\subfigure[][$k$NN graph]{\includegraphics[width=0.48\linewidth]{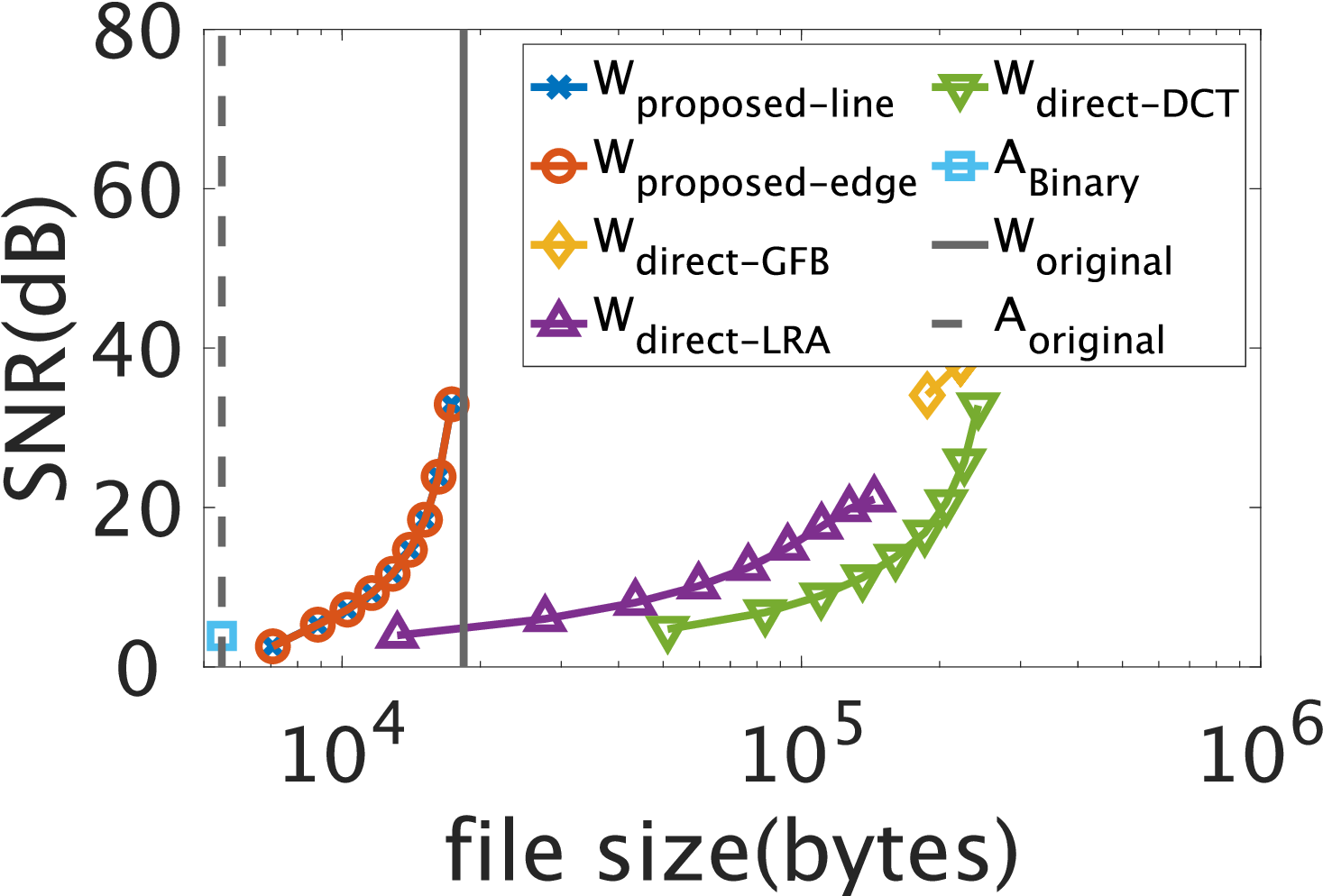}}
\subfigure[][Erd\H{o}s--R\'{e}nyi model]{\includegraphics[width=0.48\linewidth]{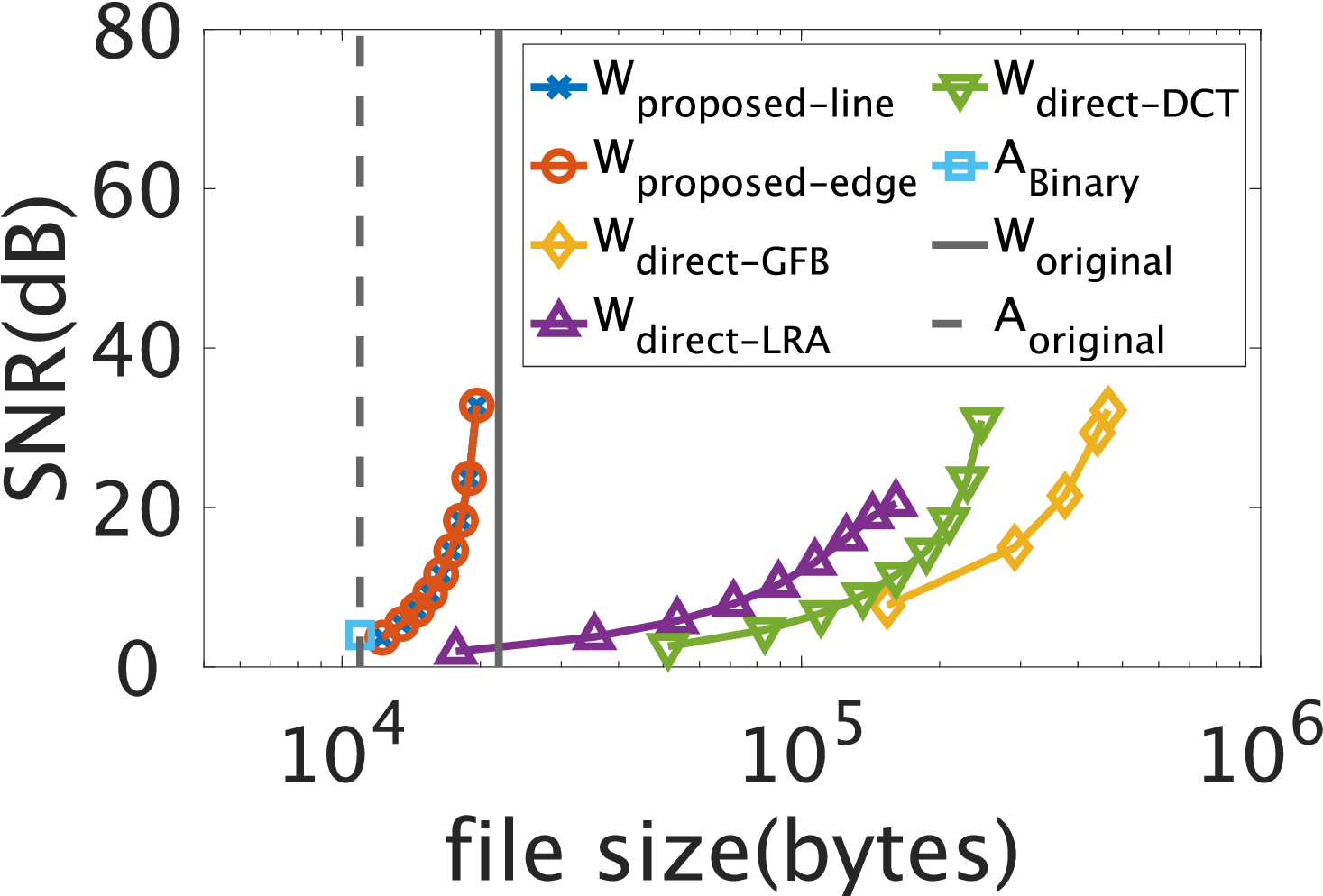}}
\caption{Performance comparison of SNRs of the reconstruction error in dB.  The horizontal axis represents the file size (bytes) when the compressed matrix or vector is stored. The black solid and dotted lines are the file sizes obtained when losslessly compressing the weighted and binary adjacency matrices, respectively.}
\label{fig:exp1}
\end{figure}

\begin{figure}[t]
\centering
\subfigure[][Sensor graph]{\includegraphics[width=0.48\linewidth]{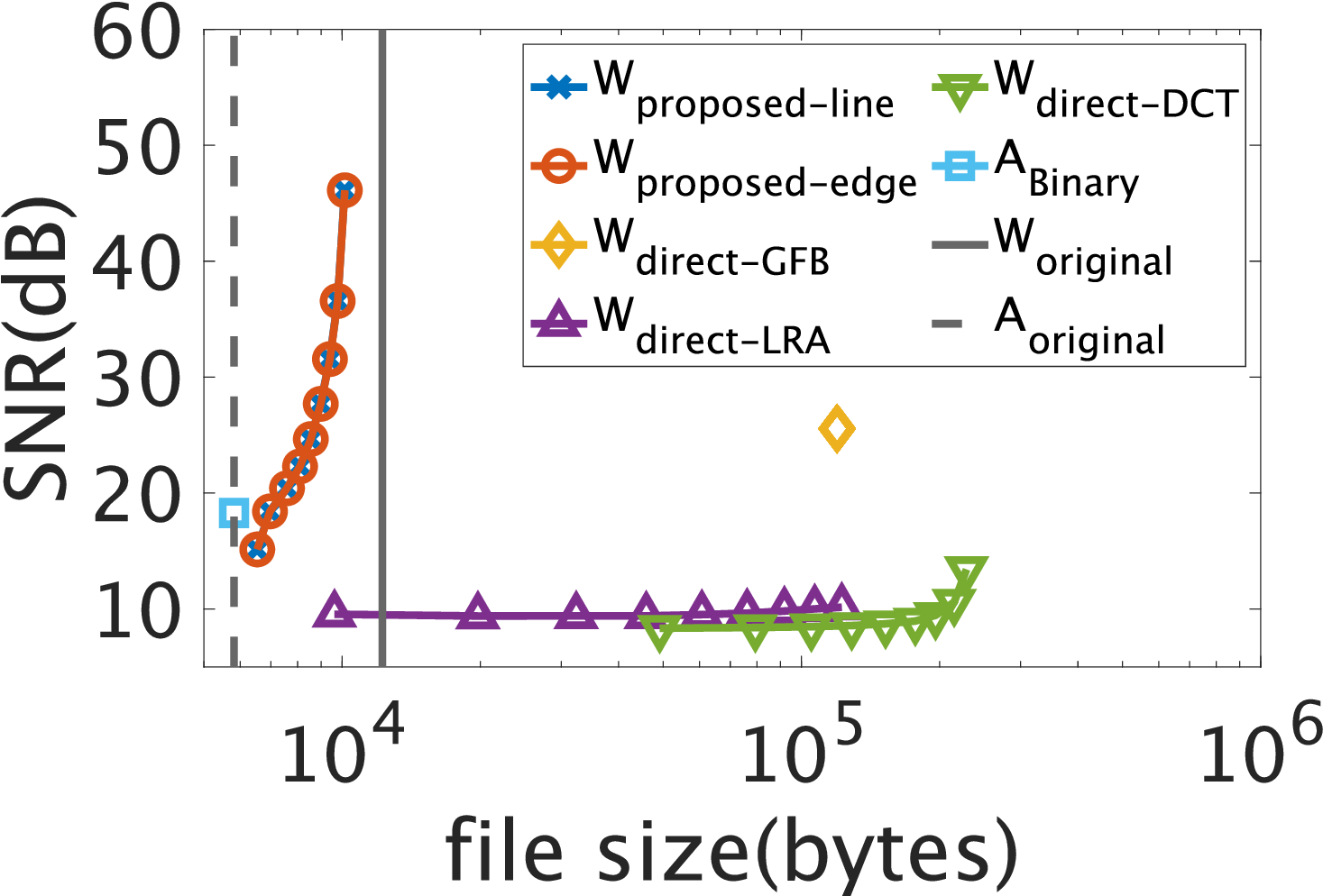}}
\subfigure[][Community graph]{\includegraphics[width=0.48\linewidth]{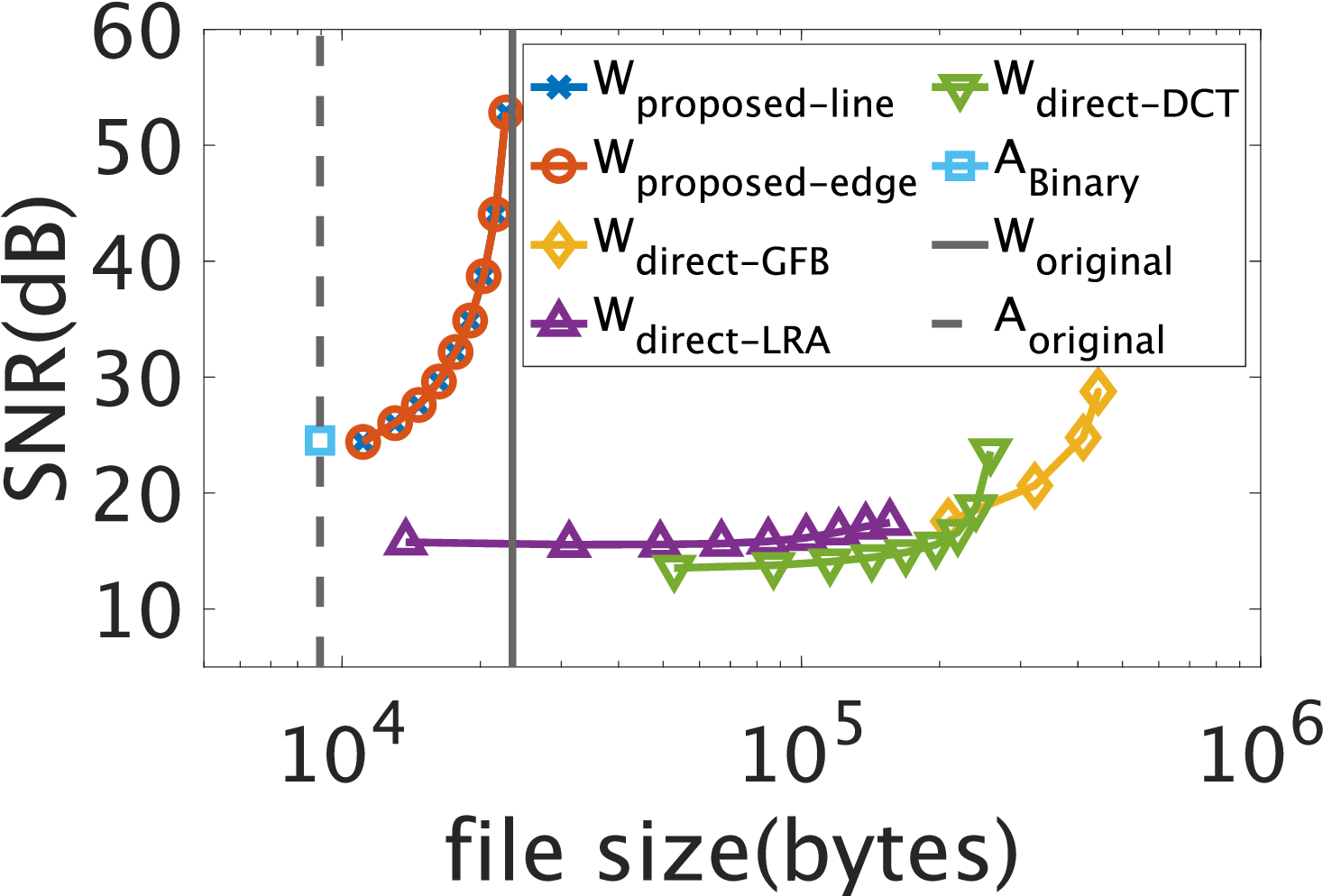}}\\
\subfigure[][$k$NN graph]{\includegraphics[width=0.48\linewidth]{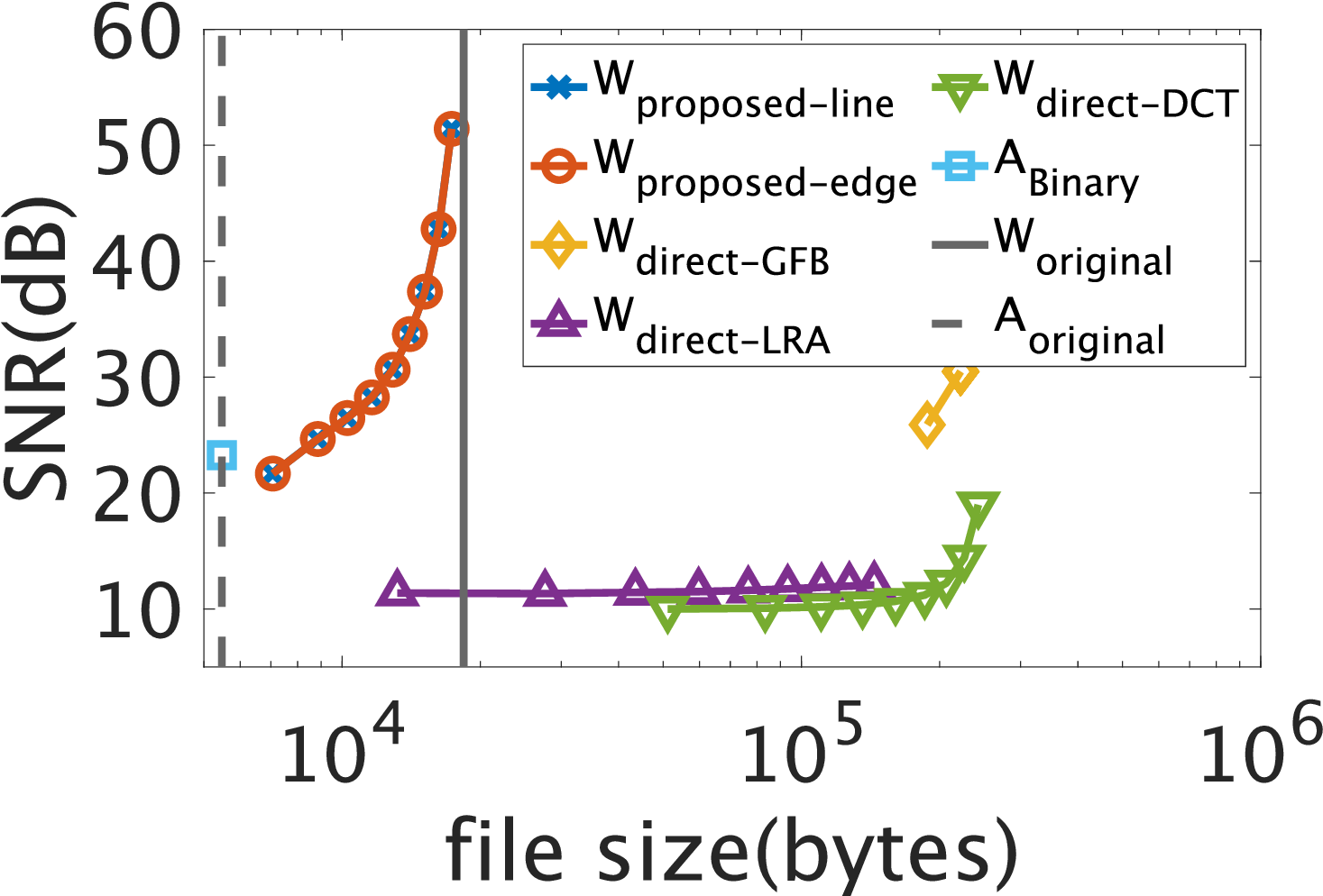}}
\subfigure[][Erd\H{o}s--R\'{e}nyi model]{\includegraphics[width=0.48\linewidth]{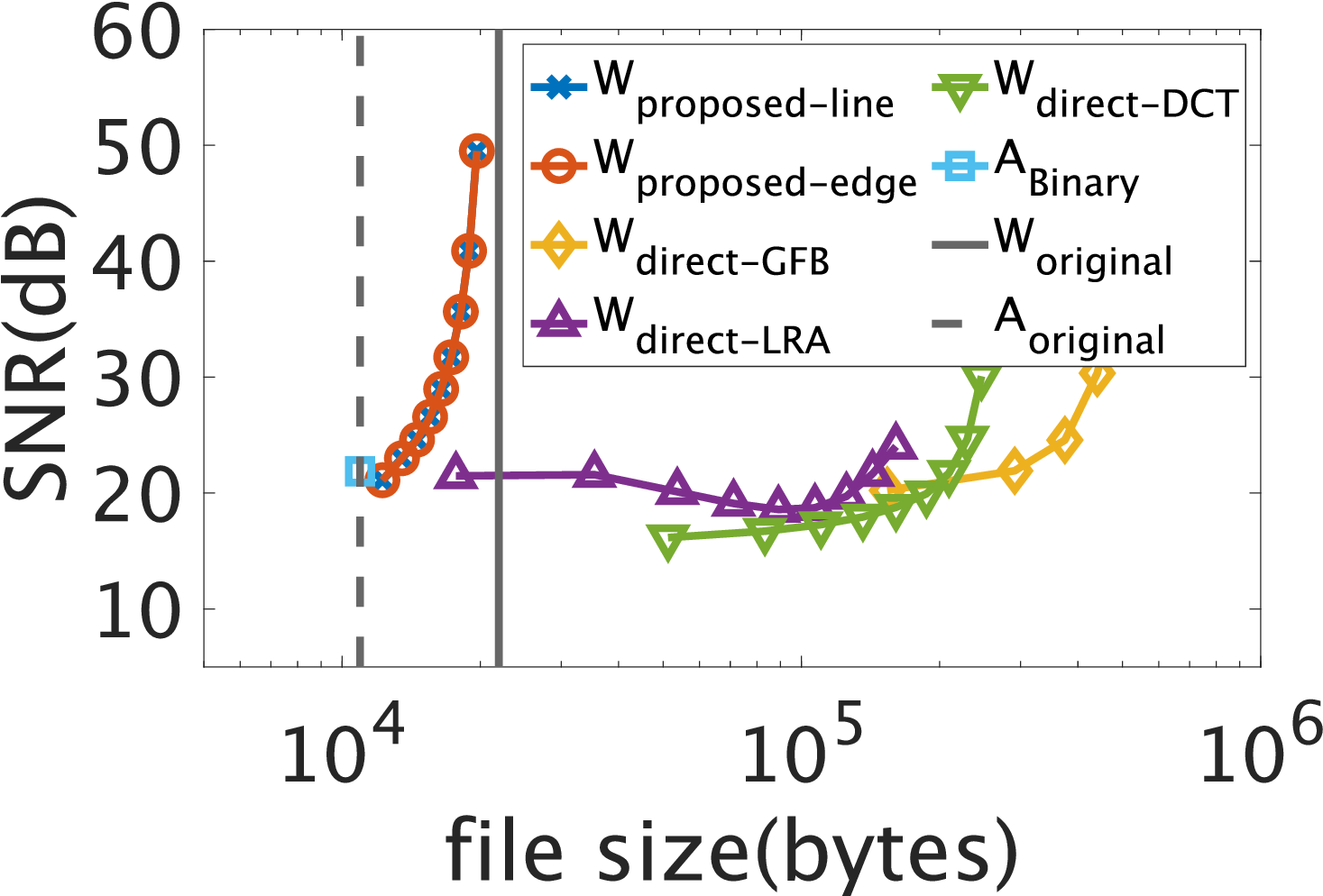}}
\caption{Performance comparison of SNRs of the diffused signals in dB. The horizontal axis represents the file size (bytes) when the compressed matrix or vector is stored. The black solid and dotted lines are the file sizes obtained when losslessly compressing the weighted and binary adjacency matrices, respectively.}
\label{fig:exp2}
\end{figure}

The accuracy of compression is compared by three metrics:

\noindent
\textbf{1) Reconstruction error}: Reconstruction accuracy is measured by signal-to-noise ratio (SNR) against the compressed file size.
We simply store the encoded coefficients with MATLAB functions and measure the file sizes to calculate the bitrates.

\noindent
\textbf{2) SNRs of diffused signals}:
As the filtered signal accuracy, SNR is measured between the diffused signals on the original and compressed graphs.
If the compressed graph preserves the original structure, the diffused signal values will be similar to those diffused on the original.
The diffused signal on the graph is represented as $\yv := \Um h(\Lambdam) \Um^\top \xv$, where $\xv \in \{0, 1\}^N$ is the input signal and $h(\Lm)$ is the graph lowpass filter.
We set $h(\lambda) = e^{-5\lambda}$.
For $\xv$, we randomly select 100 elements such that $[\xv]_i = 1$ and $0$ otherwise.

\noindent
\textbf{3) Cluster consistency before and after compression}:
Cluster consistency $C$ is represented as $C = \frac{1}{N}\sum_{i=0}^{N-1} s_i$, where $s_i$ is $1$ when the cluster assigned to node $i$ after compression is the same as that of the original graph and $0$ otherwise.
We utilize a spectral clustering method \cite{ng2001a} for the experiment.

\begin{figure}[t]
\centering
\subfigure[][Community graph]{\includegraphics[width=0.48\linewidth]{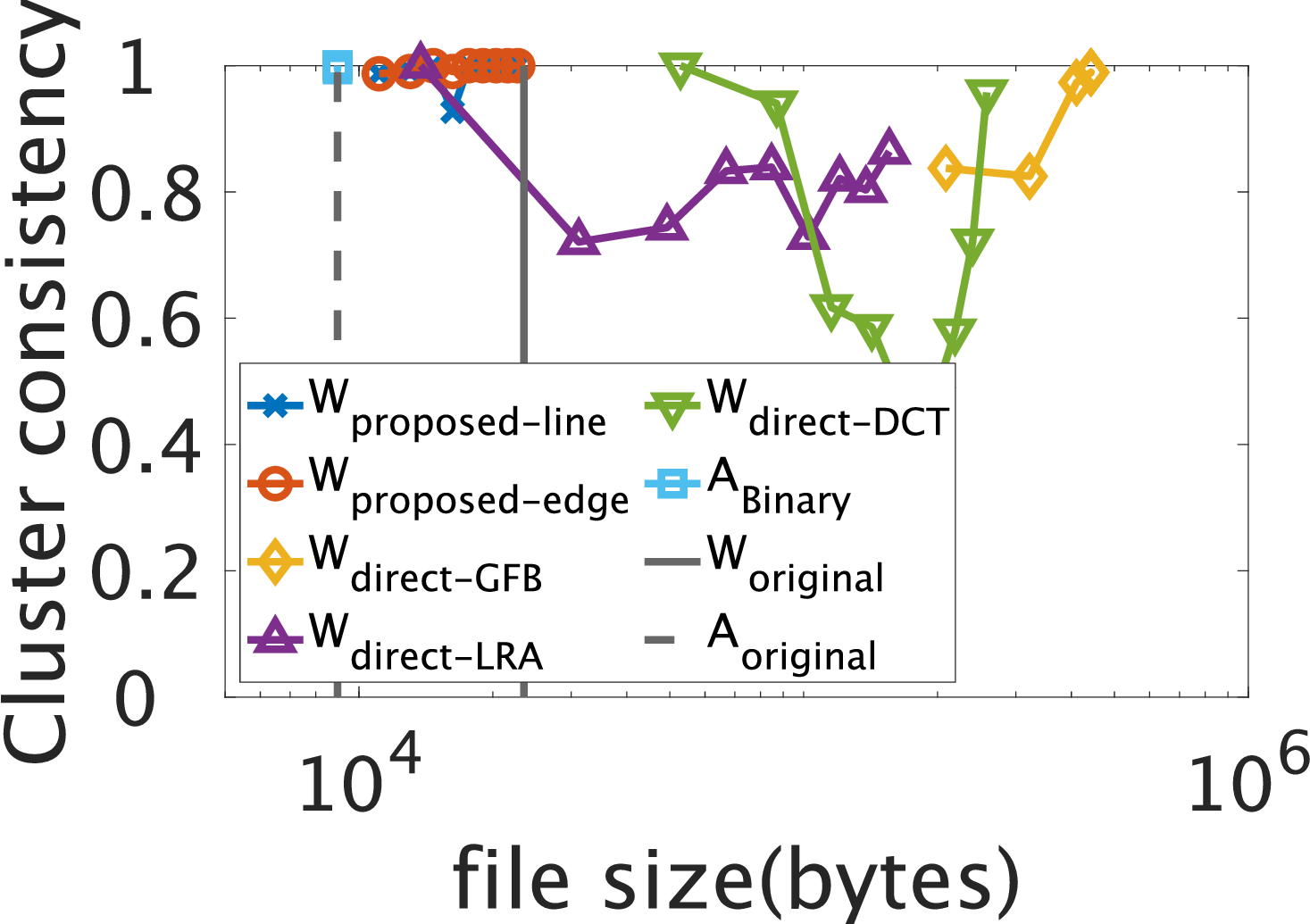}}
\subfigure[][$k$NN graph]{\includegraphics[width=0.48\linewidth]{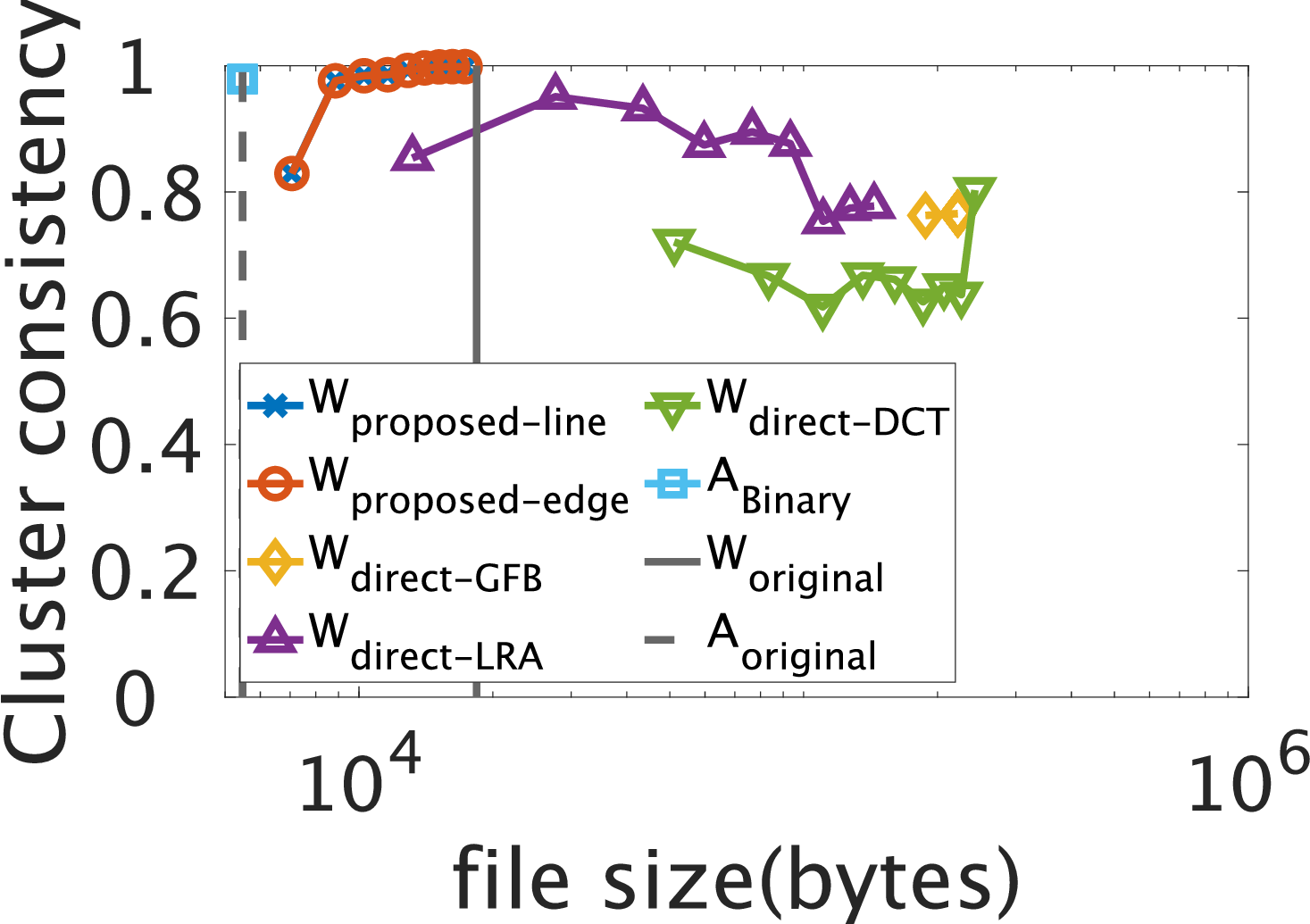}}
\caption{Performance comparison of the cluster consistency. 
The horizontal axis represents the file size (bytes) when the compressed matrix or vector is stored. The black solid and dotted lines are the file sizes obtained when losslessly compressing the weighted and binary adjacency matrices, respectively.}
\label{fig:exp3}
\end{figure}

\subsection{Results}
Fig. \ref{fig:exp1} compares the reconstruction errors.
It can be seen that the proposed methods have better reconstruction accuracy than the existing methods.
This is because our method can compress the vectorized edge weights having $|\Ec|$ elements while the other ones treat adjacency matrices directly whose number of elements is $N^2$.
The two proposed methods performed similarly in this experiment despite utilizing distinct graph Laplacians.
This similarity could stem from their construction methods: The line graph Laplacian is designed for undirected graphs, while the edge Laplacian is mainly for the directed ones. 
Thus, in our experiments on undirected graphs, the compression performance of both methods is similar.

Fig. \ref{fig:exp2} shows the SNRs of the diffused signals.
It can be seen that the proposed method presents higher SNRs than the other methods.
The proposed method, which utilizes the weighted adjacency matrix, significantly improves the SNRs compared to the lossless compression of the binary adjacency matrix.
This suggests that edge weights could have a high impact on signal filtering and the low reconstruction error of the weight information of the proposed method contributes to it.

Fig. \ref{fig:exp3} shows the consistency of the assigned clusters by spectral clustering before and after compression.
As seen from Fig. \ref{fig:exp3}, the proposed method is more consistent than other alternative methods.
The alternative methods also have a non-monotonic behavior.
The preservation of topological information and low reconstruction errors for edge weights are two possible reasons for such a performance difference.

\section{Conclusion}
This paper proposes a lossy compression framework for weighted adjacency matrices by graph filter banks.
In our method, the binary adjacency information and edge weights of the graph are compressed losslessly and lossily, respectively.
The original graph is transformed into a line graph for lossy compression of edge weights.
Edge weights are then transformed as a graph signal on the line graph.
Experimental results show the effectiveness of the proposed method against several direct compression approaches.
\bibliographystyle{IEEEtran}
\bibliography{IEEEabrv, ICASSP2024a}

\end{document}